\documentstyle[12pt]{article}

\begin{document}
\bibliographystyle{unsrt}

\begin{flushright} UMD-PP-95-93

\today
\end{flushright}

\vspace{0mm}

\begin{center}
{\Large \bf   Intermediate Scales in SUSY SO(10), b-$\tau$ Unification,
and Hot Dark Matter Neutrinos\footnote{Work supported by a grant from the
National Science Foundation}}\\ [6mm]
\vspace{9mm}

{\bf Dae-Gyu Lee and R. N. Mohapatra}\\
{\it Department of Physics, University of Maryland\\  College Park,
Maryland 20742}\\ [4mm]

\vspace{10mm}

\end {center}

\begin{abstract}
Considerations of massive neutrinos, baryogenesis as well as
fermion mass textures in the grand unified theory framework provide
strong motivations for supersymmetric(SUSY) SO(10) as the theory
beyond the standard model. If one wants to simultaneously
solve the strong CP problem via the Peccei-Quinn mechanism,
the most natural way to implement it within the framework
of the SUSY SO(10) model is to have an intermediate scale ($v_{BL}$)
(corresponding to B-L symmetry breaking) around the invisible axion scale
of about 10$^{11}$ - 10$^{12}$ GeV. Such a scale is also desirable
if $\nu_{\tau}$ is to constitute the hot dark matter ( HDM ) of the universe.
In this paper, we discuss examples of superstring inspired
 SUSY SO(10) models with intermediate scales
that are consistent with the low energy precision measurements of
the standard model gauge couplings. The hypothesis of $b-\tau$
unification which is a successful prediction of many grand unified
theories is then required of these  models and the resulting
prediction of $b$-quark mass is used as a measure of viability of
these schemes. Detailed analysis of a model with a $v_{BL}\simeq 10^{11}$
GeV, which satisfies both the requirements of invisible axion and
$\nu_{\tau}$ as HDM is presented and shown to lead to $m_b\simeq 4.9$ GeV
in the one-loop approximation.

\end{abstract}

\newpage
\section{Introduction}
\hspace{8.8mm} There are many reasons for the recent surge of interest
in supersymmetric (SUSY) SO(10) models such as
(i) a possibility to understand the observed patterns of fermion
masses and mixings\cite{SOTM}; (ii) small
non-zero neutrino masses\cite{SEE,BM93};
(iii) a simple mechanisms for baryogenesis\cite{FY86}, etc.
The recent observation\cite{CSL} that level two compactification
of superstring models can also lead to SO(10) models  with
higher dimensional multiplets such as {\bf 45} and {\bf 54}
that can help in grand unified theory (GUT) symmetry breaking has
injected new enthusiasm to this field.

In all SO(10) models considered to date, one assumes a grand desert
scenario between the TeV scale (corresponding to the electroweak
symmetry as well as SUSY breaking) and the GUT scale, M$_U$, of order
10$^{16}$ GeV. This scenario is of course required if the known low
energy gauge couplings are to unify at M$_U$ given the particle spectrum
of the minimal supersymmetric standard model (MSSM)\cite{GCU}

There are however reasons to entertain the possibility that a GUT model
such as SO(10) should have an intermediate scale around
10$^{11}$ - 10$^{12}$ GeV corresponding to B-L symmetry breaking.
The first one has to do with solving the strong CP problem via
the Peccei-Quinn mechanism\cite{PQ}. Since cosmological constraints
require the PQ-symmetry breaking scale, $v_{PQ}$, around
10$^{11}$ - 10$^{12}$ GeV, it will fit naturally into an SO(10) model
if a gauge subgroup of SO(10) such as SU(2)$_R \times$ U(1)$_{B-L}$
breaks around the same intermediate scale, {\em i.e.},
$v_{B-L} \sim v_{PQ}$.
Another reason that such model may be of interest has to do with
the possibility that the tau neutrino with a mass of few electron
volts may constitute the hot dark matter (HDM) component of the universe
needed to fit the observations on the large scale structure
with the successful big bang picture\cite{SHAFER} . In the see-saw
mechanism for neutrino masses predicted by a class of $SO(10)$ models,
this requires that there must be an intermediate scale corresponding
to the $B-L$ symmetry breaking around $10^{11}- 10^{12}$ GeV, which
is of the same order as required for the invisible axion scenario.
The fact that such an SO(10) scenario emerges naturally
in non-supersymmetric context has been known for some time\cite{RS83}.

As far as the neutrino mass alone is concerned,
 one could argue that an eV range mass for the tau neutrino
could be obtained in the
grand desert type $SO(10)$ models by  judicious "dialing" down
of the Yukawa coupling of the ${\overline {\bf 126}}$ coupling to
the matter spinors. This would of course not accomodate the invisible
axion solution to the strong CP-problem. Moreover,
in two interesting recent papers\cite{VS},
it has been noted that at least for the small $tan\beta$ case,
this alternative may run into trouble with the hypothesis of
 bottom-tau mass unification\cite{BEGN},
 which is another successful prediction of grand unified
theories.  Of course one could
abandon the $b-\tau$ unification hypothesis as in $SO(10)$ models
which contain$ {\bf 126}+{\overline {\bf 126}}$ multiplets (e.g.
see Ref.\cite{BM93} ) or one could consider a large
$tan\beta$ scenario. But if we insist on a small $tan\beta$, then
an alternative that is available is to abandon the grand desert scenario
and consider intermediate scale type $SO(10)$ models and see if it
is consistent with the $b-\tau$ unification hypothesis. In this paper,
we explore this possibility.

 We first seek simple extensions of the
minimal SUSY SO(10) model which can support intermediate scales
corresponding to B-L symmetry breaking consistent with the gauge
coupling unification. We will assume that supersymmetry is an exact
symmetry above the weak scale, as is generally believed.
 Several such models have already been
discussed in the literature\cite{SATO,DESH}. In our work, we will
assume that the particle spectrum below the Planck scale is
of the type dictated by recent level two Kac- Moody schemes\cite{CSL},
so that they contain three {\bf 16} dim. spinors corresponding to
three generations of matter fields, a number of {\bf 10},
${\bf 16}_H +{\overline {\bf 16}_H}$, {\bf 45} and {\bf 54} dimensional
fields. This is one of the respects
in which our work differs from earlier works in Ref.\cite{SATO}.
For this case, by appropriate adjustments of the particle spectrum,
we have found several new classes of intermediate scale models.
If we choose a value of the QCD coupling $\alpha_{3c}\geq .115$,
only one class of these models is singled out as preferable to the others
in the one loop approximation. Since the two loop corrections
to these results are not that drastic, we choose to do a more detailed
analysis with this model.
We then impose the additional requirement that the Yukawa couplings
corresponding to the bottom quark  and the tau lepton unify at M$_U$
and use the prediction for the bottom quark mass as an indicator of
the viability of a given scenario. We are able to find one scenario
which has $v_{BL}\simeq 10^{11}$ GeV as desired, with a prediction
for $m_b\simeq 4.9$ GeV in the one loop approximation,
which we believe is phenomenologically
acceptable\cite{neubert}. This is the main result of our paper.

We have organized this paper as follows: in Sec.~2, we discuss examples
of SUSY SO(10) models, where intermediate B-L symmetry breaking scales
can arise, consistent with gauge coupling unification; in Sec.~3,
we discuss the restrictions of b-$\tau$ Yukawa unification in these
models; in Sec.~4, we discuss how neutrino masses are understood
in this class of model; in Sec.~5, we present our concluding remarks.

\section{Gauge Coupling unification and the Intermediate Scale for
B-L Symmetry Breaking}
\hspace{8.8mm}
 It is well-known that if one assumes exact supersymmetry
above the TeV scale and the particle spectrum of the MSSM, there is no
room for an intermediate scale consistent with gauge coupling unification.
On the other hand, there exist several examples\cite{SATO,DESH} where changing
the spectrum can lead to a variety of possibilities for intermediate scales
corresponding to SU(2)$_R \times$ U(1)$_{B-L}$ symmetry breaking.
 Our goal is to
seek an intermediate scale around 10$^{10}$ - 10$^{12}$, motivated by
the allowed scale for the invisible axion solution to the strong CP
problem and an HDM $\nu_{\tau}$.
In our analysis of gauge coupling unification, we will first use
one-loop beta function to get a rough idea about the nature of
intermediate scales. We will then pick out the scenario which
has the best chance of fulfilling our requirements and do
a two loop analysis for the gauge coupling evolution to find the
more exact value of the intermediate scale. At the two-loop
level, there is a top-quark contribution to the beta-function.
In our calculation we will ignore this for simplicity of calculation.
The relevant evolution equations for the gauge couplings are:
\begin{eqnarray}
{\mbox{d} \alpha_{i} \over \mbox{d} t} =
 {\mbox{b}_{i} \over 2 \pi} \alpha_{i}^{2} + \sum_{j}
{\mbox{b}_{ij} \over 8 \pi^2} \alpha_{i}^{2} \alpha_{j}, \label{RGE}
\end{eqnarray}
where $i$=1, 2, 3, between M$_Z$ $\leq \mu \leq$ M$_R$ and denote
the U(1)$_Y$, SU(2)$_L$, SU(3)$_c$ -symmetries respectively, whereas
$i$=1, 2, 3, 4, for  M$_R$ $\leq \mu \leq$ M$_U$ and denote the
U(1)$_{B-L}$, SU(2)$_L$, SU(2)$_R$, SU(3)$_c$ -symmetries respectively.

Before presenting the detailed results, let us discuss the one loop
evolution  equations to get an idea about the nature of the
models that can support an intermediate scale.
\begin{eqnarray}
2\pi[{\alpha^{-1}_i(M_Z) - \alpha^{-1}_U(M_U) }]= b_i R + b^{\prime}_i( U -R )
\label{RGE1}
\end{eqnarray}
where we have denoted $U = ln{{M_U}\over{M_Z}}$ and $R=ln{{M_R}\over{M_Z}}$;
$b^{\prime}_2$ and $b^{\prime}_3$ stand for the values for the $SU(2)_L$
and $SU(3)_c$ beta function coefficients above the $M_R$ scale and
$b^{\prime}_1={{2}\over{5}}b_{B-L}+{{3}\over{5}}b_{2R}$. Using these one-loop
equations, several solutions were found in Ref.\cite{SATO}
where one can have $M_R\simeq 10^{11}- 10^{12}$ GeV;
 Since we are interested in solutions with similar values for $M_R$,
let us mention the two solutions found there:
\bigskip

\noindent{\it Solution A}:
 The Higgs multiplets above $M_R$ have the $U(1)_{B-L}
\times SU(2)_L\times SU(2)_R\times SU(3)_c$ (called $G_{LR}$ in what follows):
one of (2, 1, 3, 1) + (-2, 1, 3, 1); (0, 3, 1, 1); (0, 1, 1, 8) each
and two of (0, 2, 2, 1).

\bigskip

\noindent{\it Solution B}: In this case, the Higgs multiplets above $M_R$ have
$G_{LR}$ transformation properties: one of each of the following:
(2, 1, 3, 1)+ (-2, 1, 3, 1); (1, 1, 2, 1) + (-1, 1, 2, 1); (0, 2, 2, 1);
(0, 3, 1, 1) and (0, 1, 1, 8).

Note that both these solutions require the existence of the ${\bf 126}+
{\overline {\bf 126}}$ pair at the GUT scale; therefore they cannot
emerge from simple superstring models with either Kac-Moody level one or two
\cite{CSL}. We therefore seek solutions that donot involve these multiplets.
We have found six solutions using the method described in Ref\cite{RGEM}.
One of them is the one already found by
Deshpande, Keith and Rizzo\cite{DESH}. They are all characterized
two integers $(n_H,~ n_X)$, where $n_H$ represents the number of $(0,2,2,1)$
multiplets and $n_X$ represents the number of $(1,1,2,1)$+$(-1,1,2,1)$
multiplet pairs
above the scale $M_R$ . Note that these scenarios necessarily involve
D-parity breaking\cite{CMP}. Below we give the one and two loop beta
function coefficients  $b_i$ and $b_{ij}$ for different mass ranges
for these cases and in table I, we list the solutions.
\begin{eqnarray}
\begin{array}{cc}
%
%
%
%
%
%
 \mbox{(i) For $M_{SUSY} \leq \mu \leq M_R$,} \hspace*{16.5mm}
&\begin{array}{ccc}
     \left(  \begin{array}{c}
		b_1 \\ b_2 \\ b_3
         \end{array} \right)=
   & \left( \begin{array}{c}
      {33 \over 5} \\ {1} \\ -{3}
         \end{array} \right),
   & b_{ij}= \left( \begin{array}{ccc}
		   {199 \over 25}  &  {27 \over 5}  &  {88 \over 5}   \\
		   {9 \over 5}  &  {25}  &  {24}   \\
		   {11 \over 5}  &  {9}  &  {14}
                  \end{array} \right);
 \end{array}  \nonumber
\end{array} \\
\begin{array}{cc}
 \mbox{(ii) For $M_R \leq \mu \leq M_U$,} \hspace*{13mm}  & \nonumber \\
 \begin{array}{ccc}
    \left(  \begin{array}{c}
		b_1 \\ b_2 \\ b_3 \\ b_4
         \end{array} \right)=
   &\left( \begin{array}{c}
      6 + {3 \over 2} n_X \\ n_H \\ n_H + n_X \\  -3
         \end{array} \right),
    \end{array}
 & b_{ij}= \left( \begin{array}{cccc}
		   {7 + {9 \over 4} n_X}  &  {9}  &
 {9 + {9 \over 2} n_X}  &  {8}   \\
		   {3}  &  {18 +7 n_H}  &
{3 n_H}  &  {24}   \\
		   {3 + {3 \over 2} n_X}  &  {3 n_H}  &
 {18 + 7 n_H + 7 n_X}  &  {24}   \\
		   {1}  &  {9}  &  {9}  &  {14}
                  \end{array} \right).
 \end{array}
\end{eqnarray}
Note that the light (0, 2, 2, 1) Higgses
originate from 10-dim. SO(10) representation and the
light Higgs pairs of (-1, 1, 2, 1)+(1, 1, 2, 1)  originate
from 16-dim. SO(10) representation.

In order to discuss the implications of these equations,
we use the following values of $\alpha_{i} (M_Z)$\cite{PLD}
\begin{eqnarray}
\alpha^{-1}_1 (M_Z) &=& 58.97 \pm 0.05, \nonumber \\
\alpha^{-1}_2 (M_Z) &=& 29.62 \pm 0.04, \\
\alpha_{3c} (M_Z) &=& 0.120 \pm 0.013. \label{A123}
\end{eqnarray}

For a given model, the value of the intermediate scale
$M_R$ depends mainly upon
$\alpha_{3c}(M_Z)$. We display this dependence in Fig. 1 using
one-loop renormalization equations, where only the mean
values of $\alpha_i(M_Z)$ ($i$=1,2) are used.
{}From Figure~1, it is clear that only for the two models (V) and (VI)
one can have intermediate scales around 10$^{11}$ GeV,
in which we are interested.
For those two models, we give the two-loop running of
the three gauge couplings
from $ M_{SUSY}\approx m_t$ scale to $M_U$, in Figure~2 and 3.
These two-loop results are little changed from the one-loop results.
We also wish to note that between $M_Z$ and $m_t$, we use the standard
model beta functions and then treat the top mass as a threshold
correction\cite{POLONSKY}.
We have not included any other thershold corrections in our calculations.

\bigskip

\section{Constraint of b-$\tau$ Unification}
\hspace{8.8mm}

In this section, we will explore whether it is possible
to achieve b-$\tau$ mass unification in the class of intermediate scale
scenarios discussed in the previous section. Let us first give
a qualitative overview of the issues involved here. It is well-known
that Yukawa couplings for the bottom quark ($h_b$) and the
tau lepton ($h_{\tau}$) evolve
differently above the weak scale. There are two classes of diagrams
that control their evolution: (i) one class involving the virtual
gauge bosons and (ii) a second class involving virtual Higgs bosons.
We will keep their effects up to one-loop for all the quarks
and display the equations in terms of
 $Y_i\equiv {{h^2_i}\over{4\pi}}$, where $i~=~t, b, \tau$:

\begin{eqnarray}
{d Y_t \over d t} &=& {Y_t \over 2 \pi} \left[ 6 Y_t+Y_b
-\sum_{i} c^{(t)}_i \alpha_i \right],  \label{htS} \\
{d Y_b \over d t} &=& {Y_b \over 2 \pi} \left[ Y_t+6 Y_b+Y_{\tau}
-\sum_{i} c^{(b)}_i \alpha_i \right],  \label{hbS} \\
{d Y_{\tau} \over d t} &=& {Y_{\tau} \over 2\pi} \left[ 4 Y_{\tau}
+ 3 Y_b -\sum_{i} c^{(\tau)}_i \alpha_i \right],  \label{hrS}
\end{eqnarray}
where $i$ = 1, 2, 3 denote U(1)$_Y$, SU(2)$_L$, and SU(3)$_C$ respectively,
and $c^{(t)}_i$ = (13/15, 3, 16/3); $c^{(b)}_i$ = (7/15, 3, 16/3);
$c^{(\tau)}_i$= (9/5, 3, 0). We also note that we have redefined the
U(1)$_Y$ gauge coupling so that the new U(1)$_Y$-charges, $\tilde{Y}$ are
given by $\tilde{Y}=\sqrt{3 \over 5} Y$, $Y$ being the canonically assigned
U(1)$_Y$-charge.

It is well-known \cite{BBO} that if all Yukawa couplings were small,
then simple closed form solutions relating the Yukawa couplings at different
mass scales can be written down. In particular one can obtain
\begin{eqnarray}
{h_b (M_Z) \over h_{\tau} (M_Z)} &=& {h_b (M_U) \over h_{\tau} (M_U)}
{\left( {\alpha_1 (M_Z) \over \alpha_1 (M_U)} \right)}^{-{
c^{(b)}_1 - c^{(\tau)}_1 \over 2 b_1}}
{\left( {\alpha_{3c}(M_Z) \over \alpha_{3c} (M_U)} \right)}^{-{c^{(b)}_3
\over 2 b_3}} \equiv {{h_b(M_U)}\over {h_{\tau}(M_U)}} R^U_g. \label{brZ}
\end{eqnarray}

The b-$\tau$ unification scenario implies $h_b (M_U) = h_{\tau} (M_U)$.
If we ignore the effects of the U(1)$_Y$-coupling, we can easily see
that since $\alpha_3 (M_Z) > \alpha_3 (M_U)$ and $b_3 < 0$,
$h_b (M_Z) > h_{\tau} (M_Z)$ as needed.
Numerically, evaluating the right-hand side of Eq.~(\ref{brZ}), one finds
that $h_b (M_Z) / h_{\tau} (M_Z)= {m_b \over m_{\tau}} (M_Z)
\simeq R^U_g$
exceeds the experimental value (denoted by $R_{expt}$)
 by about 30\%.\footnote{We remind
the reader that in extrapolating from $m_b(m_b)$ and
$m_\tau (m_\tau)$ to their values at $M_Z$, we have used the three-loop
QCD and one-loop QED beta function effects\cite{QCED}.}
Fortunately, experimental indications of a large top quark mass
already implies that $h_t$ effects cannot be ignored in the evolution of
$h_b$; Moreover,  for large $tan \beta$ (where $tan \beta=
v_u /v_d$ with $v_u= <H^0_u>$ and  $v_d = <H^0_d>$ for MSSM),
effects of $h_b$ and $h_\tau$ cannot be ignored either.
This restores agreement between GUT scale b-$\tau$ unification and
known masses of the b and $\tau$. In a sense, given the free parameter
$tan \beta$, b-$\tau$ unification in a simple grand desert type GUT
model is not very constraining and an arbitrary choice of $h_t$ and $tan\beta$
helps in making $b-\tau$ unification an experimental success.
In fact for the case of small $tan\beta$, this can be seen explicitly
from the following formula\cite{VS}:
\begin{eqnarray}
R_{b/\tau}(M_Z)= R^U_{g}\times e^{-A_U}, \label{RBT}
\end{eqnarray}
where
\begin{eqnarray}
A_U={{1}\over{16\pi^2}}\int_{t_Z}^{t_U} h^2_t(t) dt, \label{RBT1}
\end{eqnarray}
It is clear that by adjusting $h_t$, the value of $A_U$ can be made bigger
than one producing the desired suppression.

Suppose that  we now go beyond the grand desert scenario and
require, as we do here, that there is a gauge intermediate scale, $M_R$
corresponding to SU(2)$_R \times$ U(1)$_{B-L}$ breaking.
Several new features emerge:

\noindent(i) The nature of the Yukawa couplings and their
evolutions change due to the presence of new symmetries and new
particles above $M_R$ .

\noindent(ii) The evolution of gauge couplings change
due to the same reasons.

\noindent(iii) Finally between $M_R$ and $M_U$,
$b-\tau$ mass unification implies that there is cancellation between
$h_t$ and $h_N$ (where $h_N$ is Dirac type coupling of the neutrinos)
in the evolution equation for $h_b$ so that the
correct experimental value is not guaranteed merely
by the choice of $h_t$\cite{VS}.

In this case, one can write:
\begin{eqnarray}
R_{b/\tau}(M_Z)= R^I_g.R^{IU}_g.e^{-A_I-A_{IU}}
\label{btauint}
\end{eqnarray}
where $R^I_g$ and $R^{IU}_g$ represent respectively the gauge contributions
between $M_Z$ and $M_R$ and between $M_R$ and $M_U$ and similarly
the $A_I$ and $A_{IU}$ represent the Yukawa coupling contributions in
the two ranges. The expression for $A_I$ is easily obtained from Eq.~(10)
by replacing the upper limit by $t_R$; in order to get the expression for
$A_{IU}$, we need the evolution equations for the Yukawa couplings above
$M_R$ which are given below.

 Note that in contrast with the case discussed
in\cite{VS}, where there was no gauge intermediate scale,
we have now a new possibilities to bring $R_{b/\tau}(M_Z)$
into agreement with data. The simplest way is to assume that above the
intermediate scale $M_R$, the QCD beta function coefficient $b_3$ becomes
zero, so that $\alpha_{3c}$ becomes flat; then $R^{IU}_g$ is given purely
by the $U(1)_Y$ evolution and one has roughly $R^{IU}_g\approx 1$ and
the value of $R_{b/\tau}(M_Z)$ is determined by the value of
$\alpha_{3c}(M_R)$ which is always bigger than $\alpha_{3c}(M_U)$.
This leads to $R^I_g < R^U_g$ which without any help from the top
Yukawa coupling can lead to  agreement with $R_{expt}$.
It is easy to see that this behaviour can occur if there is a color octet
with mass at $M_R$. It however turns out that in simple string inspired
models, it is difficult to get such a light octet at $M_R$. We will therefore
have to be content with the situation where there are no color fields at
$M_R$ (see table I) and see numerically what the prediction for
$R_{b/\tau}(M_Z) $ is.

Another point that is important for our discussion is the embedding
of the MSSM Higgs doublets into the GUT multiplets.
In the case of a single {\bf 10} SO(10) model (in the absence of any
{\bf 126}+{$\overline{\bf 126}$} multiplets) both
 the MSSM Higgs multiplets
are part of this {\bf 10} multiplet which leads to complete
Yukawa unification ({\em i.e.}, $h_t = h_b = h_{\tau}$ at $M_U$)\cite{ANANTH}.
While such models are quite elegant, a complete understanding of
why $tan \beta$ is large in this case becomes difficult.
We will therefore focus on the simple class of models where there are
two {\bf 10}-Higgses at the GUT scale.
In general, the Higgs doublets $H_u$, $H_d$ of MSSM would be some
linear combinations of the doublets in {\bf 10}'s. But we will adopt
a simple doublet-triplet splitting pattern such that for $\mu < M_R$,
$H_u$ and $H_d$ of MSSM arise from different {\bf 10}'s.
That such a situation is possible has been shown in Ref.~\cite{BMU}.
This assumption though not crucial for our conclusions helps to
simplify our calculations.

In the case where only one $\phi$(0,2,2,1) exists
in the intermediate scale, the Yukawa sector of the Largrangian of
the models is given by
\begin{eqnarray}
\mbox{$\cal{L}$}_Y = h_Q Q^T \tau_2 \phi Q^c +
 h_L L^T \tau_2 \phi L^c. \label{LY1}
\end{eqnarray}
The equations used for Yukawa couplings are
\begin{eqnarray}
{d Y_Q \over d t} &=& {Y_Q \over 2 \pi} \left[ 7 Y_Q + Y_L
-\sum_{i} c^{(Q)}_i \alpha_i \right],  \label{YQ1} \\
{d Y_L \over d t} &=& {Y_L \over 2 \pi} \left[ 3 Y_Q + 5 Y_L
-\sum_{i} c^{(L)}_i \alpha_i \right],  \label{YL1}
\end{eqnarray}
where $Y_Q = {h^2_Q \over 4 \pi}$ and $Y_L = {h^2_L \over 4 \pi}$;
$i$ = 1, 2, 3, 4 denote U(1)$_{B-L}$, SU(2)$_L$, SU(2)$_R$,
and SU(3)$_C$ respectively;
$c^{(Q)}_i$ = (1/6,3,3,16/3); $c^{(L)}_i$ = (3/2,3,3,0).

For reasons stated above, we will be interested in the case where
two (0,2,2,1)-Higgses appear at $M_R$. Then the Yukawa sector
of the Lagrangian at $M_R$ is then given by:
\begin{eqnarray}
\mbox{$\cal{L}$}_Y =
h_{Q_1} Q^T \tau_2 \phi_1 Q^c +
h_{Q_2} Q^T \tau_2 \phi_2 Q^c +
h_{L_1} L^T \tau_2 \phi_1 L^c +
h_{L_2} L^T \tau_2 \phi_2 L^c. \label{LY2}
\end{eqnarray}
The corresponding equations for Yukawa coupling evolution are:
\begin{eqnarray}
{d Y_{Q_1} \over d t} &=& {Y_{Q_1} \over 2 \pi} \left[
7 Y_{Q_1} + 4 Y_{Q_2} + Y_{L_1}
-\sum_{i} c^{(Q_1)}_i \alpha_i \right],  \label{YQ11} \\
{d Y_{Q_2} \over d t} &=& {Y_{Q_2} \over 2 \pi} \left[
4 Y_{Q_1} + 7 Y_{Q_2} + Y_{L_2}
-\sum_{i} c^{(Q_2)}_i \alpha_i \right],  \label{YQ12} \\
{d Y_{L_1} \over d t} &=& {Y_{L_1} \over 2 \pi} \left[
3 Y_{Q_1} + 5 Y_{L_1} + 4 Y_{L_2}
-\sum_{i} c^{(L_1)}_i \alpha_i \right],  \label{YL11}  \\
{d Y_{L_2} \over d t} &=& {Y_{L_2} \over 2 \pi} \left[
3 Y_{Q_2} + 4 Y_{L_1} + 5 Y_{L_2}
-\sum_{i} c^{(L_2)}_i \alpha_i \right].  \label{YL12}
\end{eqnarray}
For the models considered in this paper,
$c^{(Q_1)}_i=c^{(Q_2)}_i=c^{(Q)}_i$ and $c^{(L_1)}_i=c^{(L_2)}_i=c^{(Q)}_i$.
We assume that the MSSM Higgs doublets $H_u$ and $H_d$ are embedded
in $\phi_1$ and $\phi_2$ respectively.

These equations are supplemented by the one-loop evolution equations
for the gauge couplings already described in the previous
section\footnote{ Note that in the above evolution equations
only the contributions from the {\bf 10} Higgs couplings to matter
spinors are present. For models where $\overline {\bf 126}$
contributions exist, see Ref.~\cite{BB} }.
We can now write down the expression for $A_{IU}$ in Eq.~(11):
\begin{eqnarray}
A_{IU}= {{1}\over{16\pi^2}}\int^{t_U}_{t_R}[4(h^2_{Q_1}(t)-h^2_{L_1}(t))]dt
\label{AIU}
\end{eqnarray}

In writing the above equation, we have ignored the bottom and tau
Yukawa coupling effects which come from $h_{Q_2}$ and $h_{L_2}$.
We wish to point out that there is an extra factor of four in the
exponent in the above equation relative to the same equation in the
grand desert scenario\cite{VS}, which will tend to magnify our contribution
somewhat. To see the effect numerically,
we follow the procedure given below:
For a given model, using mean values for $\alpha_{1Y}(M_Z)$ and
$\alpha_{2L}(M_Z)$, choose an intermediate scale which yields
$\alpha_{3c}(M_Z)$ in the input ranges of $\alpha_{3c}(M_Z)$.
This also fixes $\alpha_U(M_U)$ as well as $M_U$.
A given $tan \beta$ together with the experimental values for
top quark and $\tau$ masses can determine the initial values for
the Yukawa couplings, $Y_{Q_i}(M_U)$ and $Y_{L_i}(M_U)$.
Then the Yukawa and gauge coupling constants are numerically
extrapolated, and the bottom quark mass is predicted.

We have scanned a large region in the $tan\beta$-$M_R$ space for
small $tan\beta$ so that the effects of $h_b$ and $h_{\tau}$ can
be ignored in the Yukawa coupling evolution equations. We find that
the best case scenario which is also physically interesting from the
point of view of invisible axion and $\nu_{\tau}$ as HDM emerges
when $M_R\simeq 10^{11}$ GeV; $tan\beta\simeq 1.7$.
In this case, $h_t(M_U)\simeq 3.54$, $\alpha^{-1}_U\simeq 23.64$
and $M_U\simeq 3.43\times 10^{15}$ GeV.
The prediction
for the bottom quark mass (pole mass) is $m_b\simeq 4.9$ GeV.
All other choices of $tan\beta$ and $M_R$ lead to larger values
of $m_b$. We
realize that this value of $m_b$ may be somewhat
 on the high side but we wish to note
that we have only used the one loop equations for the Yukawa coupling
evolution and in any case such a value is strictly not ruled out\cite{neubert}.
The evolution of the Yukawa couplings in this case are
depicted in Fig.~4.

\vspace*{4mm}

\section{Tau Neutrinos as Hot Dark Matter and Intermediate Scale
SO(10) Models}
\hspace{8.8mm}

In the previous sections, we established the existence of
simple SO(10) models with intermediate B-L symmetry breaking scales
consistent with low energy data on gauge couplings with a reasonable
prediction for $m_b/m_{\tau}$ .
Let us now discuss whether such models can indeed lead to a tau
neutrino mass in the 5 to 7 electron Volt range as required for it to be the
HDM component of the universe. The reason such a discussion is called
for is the following. A notable feature of the models we have discussed is
the absence of {\bf 126}+{$\overline{\bf 126}$}
Higgs multiplets, which are needed in the implementation
of see-saw mechanism for neutrino masses. Therefore, the existence of a
B-L breaking scale around 10$^{11}$ - 10$^{12}$ GeV does not necessarily
guarantee $m_{\nu_{\tau}}$ in the several eV mass range. We will show in
this section, that it is possible to use a generalized see-saw
mechanism\cite{GSEE} such that even without the presence of
{\bf 126}+{$\overline{\bf 126}$} representations, one can get
see-saw-like formula for light neutrino masses.
To see our proposal in detail, let us denote the Higgs-like
{\bf 16}+{$\overline{\bf 16}$} multiplets by $\Psi_H \oplus
\bar{\Psi}_H$, and matter spinors by $\Psi_a$. Let us introduce
3 gauge singlet fields, $S_a$. The part of the superpotential relevant
for neutrino masses is\footnote{ These new couplings introduced are
assumed to be sufficiently small so as not to effect the evoution equations
for the Yukawa couplings.}

\begin{eqnarray}
W_{\nu} = h^{(1)}_{ab}\Psi_a \Psi_b H_1 + f_{ab} \Psi_a \bar{\Psi}_H S_b+
M_{ab} S_a S_b. \label{WNU}
\end{eqnarray}
Recall that $H_1$ is the {\bf 10}-dim. Higgs multiplet which leads to the
$H_u$-type higgs doublet of MSSM and is therefore responsible for the
Dirac mass of the neutrinos. The resulting mass matrix involving $\nu, N$,
abd $S$ is given in the basis \{$\nu_a, N_a, S_a$, ($a$=1,2,3)\} by
\begin{eqnarray}
M_{\nu} = \left( \begin{array}{ccc}
               0                &h^{(1)}v_u      &0     \\
               h^{(1)}v_u       &0               &f v_R \\
               0                &f^T v_R         &M     		      \end{array}
\right), \label{MV}
\end{eqnarray}

Note that $h^{(1)}, f$, and $M$ are 3 $\times$ 3 matrices. it is clear that
if we ignored all generation mixings then $h^{(1)}, f$, and $M$ will be
diagonal and the mass of $a$-th light Majorana neutrino will be given by
\begin{eqnarray}
m_{\nu_a} \sim {(h^{(1)}_a v_u)^2 M_a \over f^2_a v^2_R} \label{VA}
\end{eqnarray}
If we assume that $f_a v_R \sim M_a$, then the familiar see-saw formula
results and for $v_R \sim$ 10$^{11}$ to
 10$^{12}$ GeV, $m_{\nu_{\tau}}$ is in the eV
range. Let us be clear that unlike the simple 2 $\times$ 2 see-saw models,
one cannot make definite predictions for neutrino masses and mixings
due to the presence of arbitrary
singlet mass $M_a$. Also note that
the value of $f_a$ should not be too much smaller than one
 since in our
discussion in Sec.~2 and Sec.~3, we have assumed that the right-handed
neutrino contributes to renormalization group equations for
$M_N \geq v_R$.
Looking at the formula Eq.~(\ref{VA}), one might think that regardless
of the
value of $v_R$, one might get an eV range mass for $\nu_{\tau}$ by simply
adjusting $M_{N_3}$. This is however not true; if $M_a \gg f_a v_R$,
the mass of the heavy right-handed neutrino (say $N_{\tau}$) becomes
$(f_a v_R)^2/M_a$, which is much less than $f_a v_R $.
In this case, the contribution of $N_{\tau}$ to renormalization group
evolution of Yukawa couplings will start much below $v_R$, contrary to
what is assumed in the discussion of b-$\tau$ unification. Thus, our
discussion of b-$\tau$ unification essentially restricts $M_a \approx f_a v_R$;
as a result, one recovers the usual $2\times 2$ see-saw formula for
$\nu_{\tau}$ mass and a few eV $\nu_{\tau}$ goes with a $v_R\approx
10^{11}-10^{12}$ GeV.

\vspace*{4mm}

\section{Conclusions and Outlook}
\hspace{8.8mm}

To summarize, we have analyzed the possibility that supersymmetric SO(10)
models have an intermediate scale around $10^{11}$ or $10^{12}$ GeV so that
they can naturally accomodate the invisible axion mechanism to solve the
strong CP problem and also provide room for the tau neutrino to have a
mass in the range of $6$ to $7$ eV so that it can constitute the HDM
component of the universe. We have tried to stay within the constraint
of a superstring inspired particle spectrum. We have found a scenario
which has the above property. We have then analyzed whether the desirable
property of $b-\tau$ mass unification holds in this scheme in the
small $tan\beta$ region. We find the
answer to be yes provided we accept a  value for the
pole mass value for $m_b$ of $4.9$ GeV (within the framework of a one-loop
analysis). The spectrum at the intermediate
scale needed is generated without complicated fine tuning once one
realizes that the model breaks D-parity at the GUT scale due to the presence
{\bf 45} multiplet having vacuum expectation value
along the {\bf (1,1, 15 )} direction.
Finally we wish to note that the model has enough flexibility that
one can extend it to understand the fermion mass textures (using for
example the methods of Ref.\cite{BMU}).

\section*{Acknowledgement}
\hspace{8.8mm}
We like to thank V. Barger and M. Berger for some communications.
The work of D.-G. Lee has been supported also
by a  Fellowship from the University of Maryland Graduate School.

\vspace*{8mm}

\begin{center}
{\bf Table I}\\
\begin{tabular}{|c||c|} \hline
Model & $(n_H,~ n_X)$ \\ \hline
I   &   $ (1,~ 2)$  \\
II  &   $ (1,~3)$ \\
III &  $(1,~ 4)$ \\
IV  &  $(2,~3)$\\
V  &   $(2,~4)$\\
VI  &  $(2,~5)$\\ \hline
\end{tabular}
\end{center}

\noindent{\bf Table Caption}: In this table, we give the
Higgs particle contents at the scale $M_R$ that define
the different models; The symbols $(n_H , n_X )$ denote the
numbers of (0, 2, 2, 1) and (1, 1, 2, 1)+(-1, 1, 2, 1)
multiplets respectively.

\vspace*{16mm}

\noindent{\bf Figure Caption}\\

\noindent{\bf Figure 1:} The values of the intermediate mass scale $M_R$
for different scenarios as a function of the $\alpha_{3c}(M_Z)$
in the one-loop approximation. The
case II corresponds to a vertical line where all values of intermediate
scales are allowed in the one-loop approximation, since the evolution
equation becomes independent of $M_R$.\\

\noindent{\bf Figure 2}: Evolution of gauge coupling  in the two-loop
approximation for the scenario V.\\

\noindent{\bf Figure 3}: Evolution of the gauge couplings in the two-loop
approximation for the scenario VI.\\

\noindent{\bf Figure 4}: (a) Evolutions of the gauge couplings in the
one-loop approximation  and
(b) evolution of the Yukawa couplings $h_t$ (
labelled A), $h_N$ (labelled B),
and the ratio $h_b/h_{\tau}$ ( labelled C ) for
 the scenario V which is physically most interesting.

\end{document}